\documentclass[conference]{IEEEtran}
\IEEEoverridecommandlockouts

\usepackage{cite}
\usepackage{amsmath,amssymb,amsfonts}
\usepackage{algorithmic}
\usepackage{graphicx}
\usepackage{textcomp}
\usepackage{xcolor}
\usepackage{mathtools}
\usepackage{bm}
\usepackage{adjustbox}
\usepackage{tikz}
\usepackage[justification=centering]{caption}
\usetikzlibrary{fit,tikzmark,arrows.meta,calc}
\usepackage[caption=false,font=footnotesize,labelfont=sf,textfont=sf]{subfig}
\usepackage{caption}
\usepackage{gensymb}
\usepackage{graphicx}
\usepackage{array}
\usepackage{booktabs}
\usepackage{tabularx}
\graphicspath{ {./Figures/} }
\def\BibTeX{{\rm B\kern-.05em{\sc i\kern-.025em b}\kern-.08em
    T\kern-.1667em\lower.7ex\hbox{E}\kern-.125emX}}

\begin{document}

\title{Generative Models for Modeling and Synthesizing MIMO Channels in Adverse Weather Conditions}

\author{%
    \IEEEauthorblockN{%
        Vignesh Nandakumar,
        Faraz Barati,
        Brian L. Evans}
    \IEEEauthorblockA{%
        6G@UT Wireless Research Center\\
        The University of Texas at Austin, Austin, TX USA\\
        Emails: vnandakumar@utexas.edu,
        faraz.barati@utexas.edu,  bevans@ece.utexas.edu}
}

\maketitle

\begin{abstract}
The push for broader coverage in future cellular networks depends on reliable service, yet this is increasingly harder to do as we encounter more instances of extreme weather conditions. In extreme weather conditions, we have difficulty evaluating coverage due to limited access to channel measurements. In this paper, we generate channel state information (CSI) in low and moderate weather conditions to synthesize realistic MIMO CSI under adverse weather conditions. 
Our primary contributions are to (1) synthesize MIMO channel datasets incorporating three weather types, each with three intensity levels, representative of practical 5G/6G scenarios; (2) train a diffusion model conditioned on weather using channel samples obtained through conventional pilot-based estimation under low and moderate weather intensities, and subsequently use it to generate channel realizations for severe weather conditions; and (3) evaluate the downlink Bit Error Rate (BER) and Outage Probability measures using the generated channels. The results show that diffusion-based generative models provide a scalable, data-driven alternative for channel modeling in harsh environments and can generalize to severe weather conditions using only low- and moderate-intensity training data.
\end{abstract}


\section{Introduction}
Wireless communication supports a wide range of modern applications, from everyday mobile connectivity to mission-critical systems such as emergency response, autonomous vehicles, and industrial automation. As 5G networks mature and research accelerates toward 6G and beyond, the robustness of wireless systems in challenging and dynamic environments becomes increasingly important, especially for applications requiring extreme reliability and low-latency communication.

A key vulnerability arises during \emph{adverse and extreme weather conditions}. Events such as heavy rain, snowstorms, fog, and sandstorms introduce substantial degradation through increased path loss, phase distortion, scattering, and severe fading. These impairments directly translate into risks for safety-critical applications, including disrupted emergency communications and unreliable connectivity in autonomous or mobile robotic systems. While classical channel models provide accurate descriptions under nominal conditions, they rarely capture the nonlinear, non-stationary behavior of wireless channels during extreme weather. The scarcity of empirical datasets, due to the infrequency and measurement difficulty of such events, further limits the model development.

\begin{figure}[t]
    \boxed{\includegraphics[trim = 0cm 5cm 0cm 5cm,width=\columnwidth]{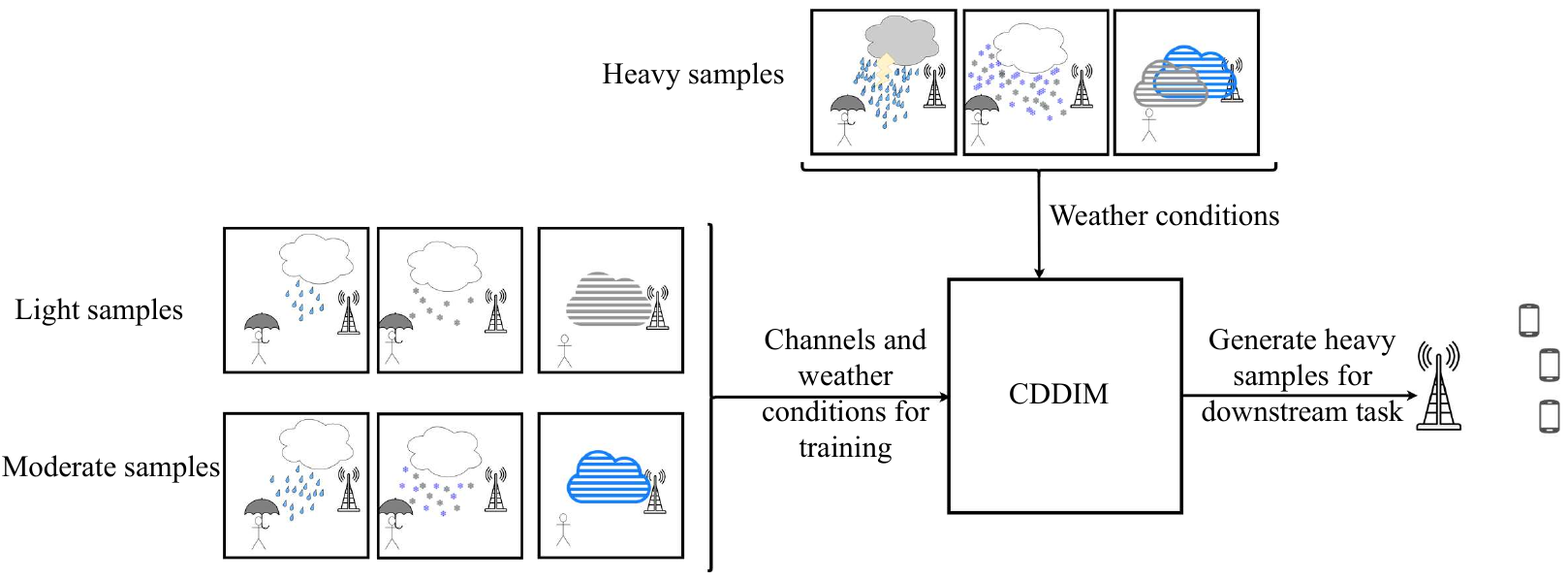}}
    \captionsetup{font=small}
    \caption{Pipeline for dataset generation and downlink evaluation.}
    \label{fig:system}
    \vspace{-3mm}
\end{figure}

To address these limitations, recent advances in \emph{AI- and data-driven wireless communication} have demonstrated significant promise \cite{vtc1,vtc2,vtc3}. Machine learning (ML) methods are increasingly being used to model complex, high-dimensional wireless channels, surpassing the capabilities of analytic models that rely on idealized assumptions such as Gaussianity or stationarity. Among ML approaches, \emph{generative models} have emerged as powerful tools for channel modeling, CSI compression, and synthetic data generation. Models such as variational autoencoders, generative adversarial networks, and denoising diffusion models (DDMs) can learn the underlying distribution of channel measurements and generate high-fidelity samples that improve signal processing tasks\cite{orekondy2022mimogan,sengupta2023diffusion,huynh2024genai}.

Generative models are particularly beneficial for scenarios with limited real-world measurements, including edge deployments, industrial internet of things (IoT) systems, and rare extreme-weather events. By synthesizing realistic channels, they support improved training of downstream tasks such as beamforming, precoding, and mobility prediction, thus enhancing resilience for 5G/6G systems. As wireless communication progresses toward AI-native architectures, environmental awareness, and robust multi-antenna operation, integrating generative modeling into the physical-layer design has the potential to substantially advance system performance.

Several studies have analyzed the impact of adverse weather on wireless propagation. Terahertz (THz) measurements under rain and snow reveal significant increases in path loss and delay spread \cite{terahertz}. Work on holographic reconfigurable intelligent surfaces (HRIS) under foggy conditions \cite{RIS} shows the sensitivity of such arrays to atmospheric disturbances, and UAV-assisted networks in extreme environments highlight the interplay between mobility, channel variability, and reliability \cite{UAV}. Rainfall-induced fading in MIMO systems has also been modeled using stochastic propagation techniques \cite{rainfall_2015}. On the generative modeling front, conditional diffusion models have been used to synthesize high-dimensional channel data \cite{cddim}, score-based models have demonstrated robustness for channel estimation \cite{score-based}, and denoising diffusion models have been applied to enhance semantic communication systems \cite{cddm}. However, existing work does not address the problem of \emph{MIMO channel synthesis under extreme weather conditions}, nor does it provide systematic evaluation strategies for generative models in this setting.

A key challenge for diffusion-based channel modeling is the lack of a known ground-truth distribution for MIMO channels, making direct evaluation of generated samples difficult. Additionally, no publicly available datasets capture multi-antenna channel behavior across varying weather intensities.

\textbf{Contributions:} To address these gaps, this paper makes the following contributions:
\begin{enumerate}
    \item We construct a synthesized MIMO channel dataset incorporating three weather types, each with three intensity levels, representative of practical 5G/6G scenarios.
    \item We adopt a conditional diffusion model and train it on weather using channel samples obtained through conventional pilot-based estimation under low and moderate weather intensities, and subsequently use it to generate channel realizations for severe weather conditions.
    \item We assess downstream \emph{Bit Error Rate (BER)} and \emph{Outage Probability} measures using generated channels based on test labels of heavy weather intensities, demonstrating relevance for signal processing and multi-antenna communication design.
\end{enumerate}
The diffusion model framework is adopted based on \cite{cddim}, but modified to be conditioned on weather based labels instead of user location.

The results indicate that diffusion models can effectively learn the relationship between weather conditions and channel characteristics, thereby enabling synthetic channel generation for scenarios in which real measurements are limited or unavailable. Furthermore, the proposed approach removes the need for explicit knowledge of the underlying ground-truth channel distribution, relying instead solely on channel samples obtained through conventional estimation methods that are readily deployable in practice, but can be replaced by field-data if available. These findings highlight the potential of AI-native channel modeling pipelines to aid system performance under severe environmental impairments.

The remainder of this paper is organized as follows. Section~II describes the system model and simulation setup. Section~III presents numerical results and analysis. Section~IV concludes the paper and outlines future research directions.

\section{Methodology}

\subsection{Weather-Dependent Attenuation Model}
Understanding how wireless channels behave under diverse weather conditions is essential for designing adaptive and robust 5G/6G systems. We focus on three common weather environments: rain, fog, and snow; and model the effects using well-established attenuation models from the ITU standards \cite{ITU_rain,ITU_fog,ITU_snow} and related literature \cite{terahertz,UAV}.

The specific attenuation $\gamma$ (dB/km) under different weather scenarios is modeled as
\begin{equation}
\label{weather_eq}
   \gamma (dB/km) =
    \begin{cases}
        kR^\beta, & \text{Rain} \\
        K_1(f,T)M, & \text{Fog} \\
        (1.023*10^{-4}\lambda_{nm} + 3.786)R^{0.72}, & \text{Snow}
    \end{cases}
\end{equation}
where $R$ is the precipitation rate (mm/h for rain, snowfall speed for wet snow), $M$ is fog water density ($g/m^3$), and $(k,\beta)$ are frequency-dependent parameters from \cite{ITU_rain}. The fog model uses
\begin{equation}
    K_1(f,T) = \frac{0.819f}{\epsilon''(1+\eta^2)}, \quad 
    \eta = \frac{2+\epsilon'}{\epsilon''},
\end{equation}
where the complex permittivity of water $(\epsilon',\epsilon'')$ is derived using temperature-dependent relaxation frequencies \cite{ITU_fog}. For snow, attenuation depends on both wavelength and snowfall intensity.

These models capture the large-scale weather-induced attenuation incorporated in our channel simulations and are directly related to the optical depth that will be briefly discussed in the next section.

\subsection{Stochastic MIMO Channel --- Weather-Induced Scattering}
Beyond large-scale attenuation, weather phenomena introduce absorption and scattering effects due to randomly distributed particles (rain droplets, snowflakes, fog aerosols). To the best of our knowledge, there are no deterministic models that accurately describe this phenomenon, and no datasets are available for the specific parameters and configuration considered in this work. Hence, we use the stochastic scattering framework in \cite{scatter}, which models a narrowband single user MIMO (SU-MIMO) channel with $N_t$ uniform linear array (ULA) transmit and $N_r$ ULA receive antennas, separated by distance $D$. We assume that, in the absence of atmospheric particles, the channel consists solely of a line-of-sight (LOS) component. When particles are present, additional scattering paths are introduced by these particles. In other words, reflectors from surrounding objects are not considered in this work. Assuming parallel ULAs with zero elevation/azimuth angles, the deterministic free-space component is
\begin{equation}
    \hat{h}_{m,n}=e^{-j\frac{2\pi D}{\lambda}}
    e^{-j(\phi_r^{(m)}+\phi_t^{(n)})}
    e^{j\frac{S(2m-1-N_r)(2n-1-N_t)}{2}},
\end{equation}
where $\lambda$ is the wavelength, and $\phi_r^{(m)}$, $\phi_t^{(n)}$, and $S$ capture array-induced phase shifts.

The complete weather-impaired MIMO channel is modeled using the Kronecker framework \cite{kermoal2002stochastic}, and is expressed as:
\begin{equation}
\label{channel}
    H = \bar{H} + \beta A^{1/2} W B^{1/2},
\end{equation}
where
\begin{itemize}
    \item $\bar{H}=\sqrt{\rho_0}\,\alpha\,\hat{H}$ is the direct (attenuated) component,
    \item $W\!\sim\!\mathcal{CN}(0,1)$ models random scattering,
    \item $A$ and $B$ encode receive/transmit spatial correlation,
    \item $\beta=\sqrt{\rho_s}\,\alpha$ scales the scattered contribution.
\end{itemize}

Here, $\alpha=\frac{\lambda}{4\pi D}$ is the large-scale propagation factor, $\rho_0=e^{-\tau_0}$ and $\rho_s=e^{-\tau_a}-e^{-\tau_0}$ depend on optical depth ($\tau_0$), absorption depth ($\tau_a$), and scattering depth ($\tau_s$). Spatial correlation matrices are
\begin{equation}
    A=\tfrac{1}{N_r}\Psi_r \odot G_r,\quad 
    B=\tfrac{1}{N_t}\Psi_t \odot G_t,
\end{equation}
where $G_r=\hat{H}\hat{H}^H$ and $G_t=\hat{H}^H\hat{H}$,  $\Psi_r =\{ \psi_r^{(m,m')}\} \in \mathbb{C}^{N_r \times N_r}$ and $\Psi_t =\{ \psi_t^{(n,n')}\} \in \mathbb{C}^{N_t \times N_t}$ are full matrices with $\psi_r^{(m,m')} = e^{-\frac{(m-m')^2d_r^2}{l_0^2}}$ and $\psi_t^{(n,n')} = e^{-\frac{(n-n')^2d_t^2}{l_0^2}}$ that encode exponential decay correlations with scattering length $l_0$ \cite{scatter}. This model captures how weather severity influences both large-scale attenuation and small-scale fading statistics.

\subsection{Diffusion-Based Generative Channel Modeling}
\begin{figure}[t]
    \includegraphics[width=0.9\columnwidth]{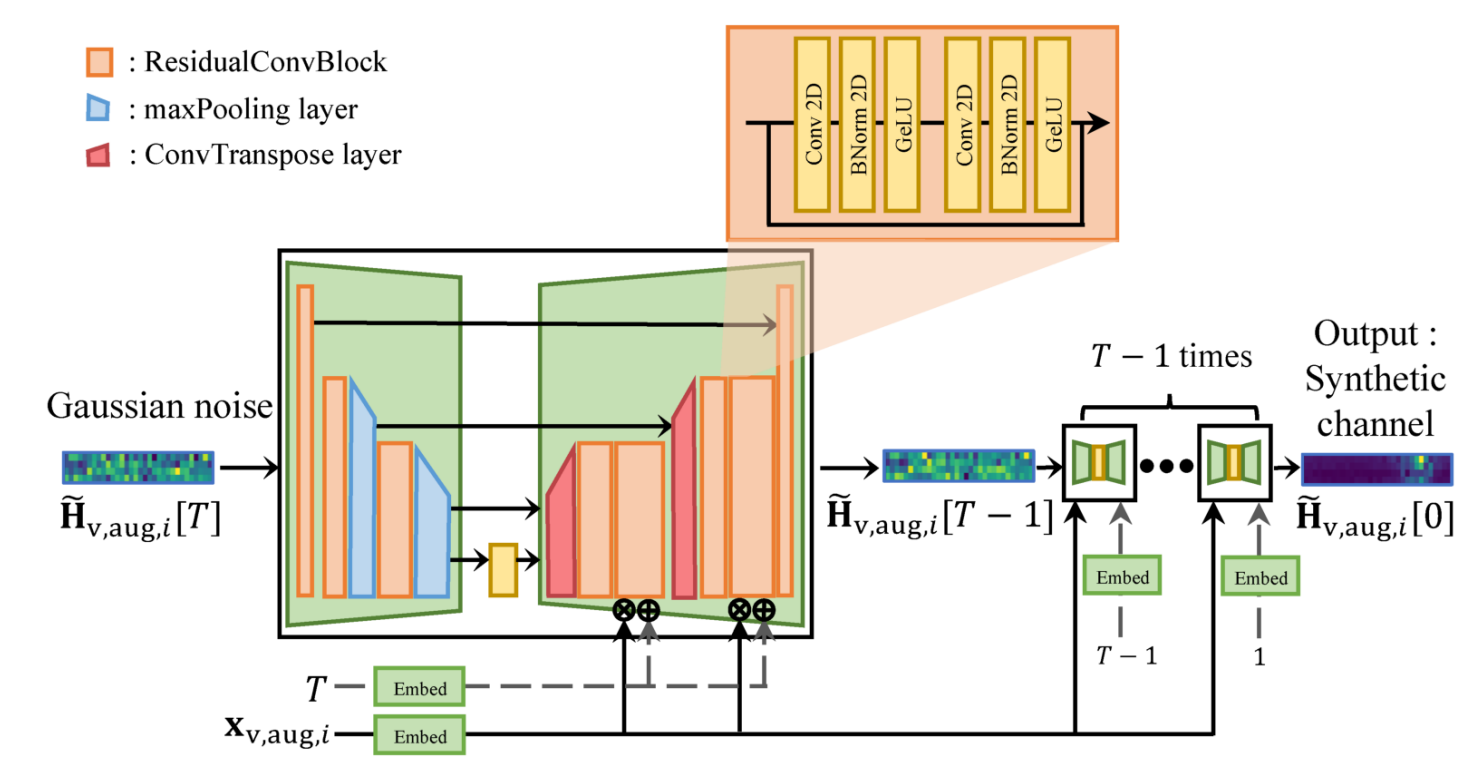}
    \captionsetup{font=small}
    \caption{Illustration of cDDIM's inference process \cite{cddim}.}
    \label{fig:inference}
\end{figure}

Diffusion models generate data by reversing a gradual noising process. In the forward direction, a clean sample is progressively corrupted through a sequence of noise-injection steps until it becomes nearly Gaussian. The generative process performs the reverse operation by iteratively denoising this Gaussian sample according to a learned score function. This reverse-time dynamics allows the model to reconstruct samples that follow the target data distribution. Given a conditioning variable $X$ (e.g., weather type and intensity), our goal is to model the conditional channel distribution $p(H|X)$. Because $p(H|X)$ is unknown and only empirical samples are available, we follow the conditional denoising diffusion implicit model (cDDIM) framework, where the inference process has been demonstrated in Figure~\ref{fig:inference}. Further details on the cDDIM system model can be found in \cite{cddim}.

Let $H[t]$ denote the noisy channel at step $t$. The score function $\nabla_{H|X}\ln p(H|X)$ is approximated using a deep neural network $S(H|X;\Theta)$ trained via denoising score matching. Because the true score is unknown, noisy observations $\tilde{H}=\zeta H+\sigma N$ with $N\sim \mathcal{N}(0,I)$ yield the practical loss
\begin{equation}
\label{loss}
    \mathcal{L}=\tfrac{1}{2}\mathbb{E}_{N}\Big\|-\sigma S(\tilde{H}|X;\Theta)-N\Big\|^2,
\end{equation}
where $\zeta$ is a constant and $\sigma$ is the variance of the noise. It has been shown that the loss mentioned in \eqref{loss} is equivalent to the ideal score-matching objective \cite{cddim, score-based}. 
During training, each clean channel realization $H_{\text{train},i}[0]$ is randomly assigned a diffusion 
step $t \sim \text{Uniform}(1,T)$ and corrupted according to the predefined 
noise schedule $\zeta[t]$, producing a perturbed sample
\begin{equation}
H_{\text{train},i}[t] = \sqrt{\bar\zeta[t]} H_{\text{train},i}[0] + \sqrt{1-\bar\zeta[t]} N[t],
\end{equation}
where $N[t] \sim \mathcal{CN}(0,1)$ captures the injected noise. The score 
network $\tilde{S}(\cdot \mid \cdot;\Theta)\triangleq -\,\sigma\, S(\cdot \mid \cdot; \Theta)$ is then updated by minimizing the Frobenius-norm discrepancy between its output and the true noise $N[t]$, enabling the model to predict the score function associated with the conditional channel distribution $p(H \mid x)$. The exact formulation for the parameter $\bar\zeta[t]$ has been given in \cite{cddim}.

Once trained, the model is used for conditional channel generation. Starting from pure Gaussian noise 
$H_{\text{aug},i}[T] \sim \mathcal{CN}(0,1)$, the sampler iteratively applies the reverse diffusion update using the learned score network and the same schedule $\zeta[t]$, progressively denoising the sample until $H_{\text{aug},i}[0]$ is obtained. The resulting realizations approximate draws from the conditional distribution $p(H \mid x)$ for any desired weather condition $x$.

\subsection{Wasserstein Distance for Statistical Evaluation}
Since the ground-truth conditional distribution $p(H|X)$ is unknown, traditional metrics such as KL divergence cannot be applied. Instead, we employ the Wasserstein distance, a geometry-aware metric widely used in WGANs \cite{wgan} and suitable for limited or empirical datasets.

Given empirical distributions $P$ and $Q$ with samples $\{X_{(i)}\}$ and $\{Y_{(i)}\}$, the Wasserstein $p$-distance is
\begin{equation}
    W(P,Q)=\left(\frac{1}{n}\sum_{i=1}^{n}\|X_{(i)}-Y_{(i)}\|^p\right)^{1/p},
\end{equation}
measuring the minimum “transport cost" between the two distributions. This allows us to quantify how closely diffusion-generated channels resemble those from the physical scattering model under unseen weather intensities.

\subsection{Downstream Evaluation: BER and Outage}
To assess practical utility, we evaluate the generated channels in a downstream communication task. Specifically, we compute the BER and outage probability under a precoding-based MIMO system and compare the results against conventional channel estimation techniques. This evaluates whether diffusion-generated samples preserve key physical-layer characteristics relevant for 5G/6G signal processing.

\section{Simulation and Results}
For our simulations, we consider rainfall, snow, and fog conditions that are categorized as light, moderate, or heavy using the parameter ranges shown in Table~\ref{table:1}, following standard references \cite{ITU_fog,rain_ref,snow_ref,fog_ref}. The temperature for fog simulations is set to $293.15K$, which is used in equation \eqref{weather_eq}.

\begin{table}[h]
\centering
\caption{Parameters for Weather Conditions}
\label{table:1}
\begin{tabular}{|l|c|c|c|}
\hline
\textbf{Condition} & \textbf{Light} & \textbf{Moderate} & \textbf{Heavy} \\ \hline
Rain rate (mm/h)        & $0$--$2.5$       & $2.5$--$7.5$        & $7.5$--$50$ \\ \hline
Snow rate (mm/h)        & $0$--$1$         & $1$--$2.5$          & $2.5$--$10$ \\ \hline
Fog density (g/m$^3$)   & $0$--$0.05$      & $0.05$--$0.2$       & $0.2$--$0.5$ \\ \hline
\end{tabular}
\end{table}

In this work, we do not vary particle-size distributions across weather intensities. Instead, the anisotropy factor is treated solely as a function of the medium type. Incorporating particle-size variability into the scattering model is left as a direction for future work. The diffusion model is conditioned on the type of the weather (rain, fog or snow), the intensity of the weather (precipitation rate) and the attenuation coefficient. Hadamard matrices were employed to construct the pilot sequences due to their orthogonality properties, which enable efficient separation of transmit antennas and improve the conditioning of the least-squares (LS) channel estimation problem. This provides low-complexity and robust channel estimation while minimizing inter-pilot interference. For channels under light and moderate weather conditions, the training samples consist of pilot-based channel estimates obtained at SNR of $16\;\mathrm{dB}$.  We assume a single-user MIMO model under millimeter-wave with a fixed distance $D$ between the basestation (BS) and user, where the center frequency ($f_c$), the number of reception antennas $N_r$, and the number of transmit antennas $N_t$ are listed in Table \ref{table:2}.

\begin{table}[h]
\centering
\caption{Simulation Parameters}
\label{table:2}
\begin{tabularx}{\columnwidth}{|c|c!{\vrule width 1.5pt}c|X|} 
\hline
\multicolumn{2}{|c!{\vrule width 1.5pt}}{\textbf{Channel Generation}} &
\multicolumn{2}{c|}{\textbf{Training}} \\
\hline
\textbf{Parameter} & \textbf{Value} & \textbf{Parameter} & \textbf{Value} \\
\hline
$N_t$ & $32$ & DDIM train epochs & $15{,}000$ \\ \hline
$N_r$ & $4$ & DDIM sampling steps & $256$ \\ \hline
$f_c$ & $28~\mathrm{GHz}$ & Batch size & $1024$ \\ \hline
$D$ & $50~\mathrm{m}$ & Number of training samples & $12{,}000$ \\ \hline
$d_r = d_t$ & $0.45\lambda$ & Number of Monte Carlo iterations & $10{,}000$ \\ \hline
\end{tabularx}
\end{table}
\subsection{Wasserstein Metric}
Although the Wasserstein distance provides a robust, geometry-aware measure of similarity between probability distributions, its absolute magnitude is difficult to interpret compared to likelihood-based metrics. Ideally, the diffusion model would be trained using different percentages of ground-truth data to identify which percentage yields the most accurate prediction. Such an analysis would clarify how data quantity influences generative performance, but repeatedly retraining large cDDIM models is computationally expensive.

As a practical alternative, we examine the evolution of the Wasserstein distance across training epochs. A clear inflection in this trajectory would indicate convergence in the model’s ability to represent the distribution of the adverse-weather channel, serving as a proxy of data sufficiency. Only for statistical analysis, we train the model once using channel samples based on \eqref{channel} (the ground truth) and once with channel samples based on pilots. Each method only uses light and moderate weather intensities to evaluate whether the model can accurately predict CSI samples in a high-intensity distribution. To provide a meaningful comparison, we also report the Wasserstein metric for the conventional pilot-based channel estimation method at a low SNR of $6\,\mathrm{dB}$, which corresponds to the expected operating point under severe weather conditions. 
\begin{figure}[h]
\vspace{-3mm}
\includegraphics[width=1.0\linewidth]{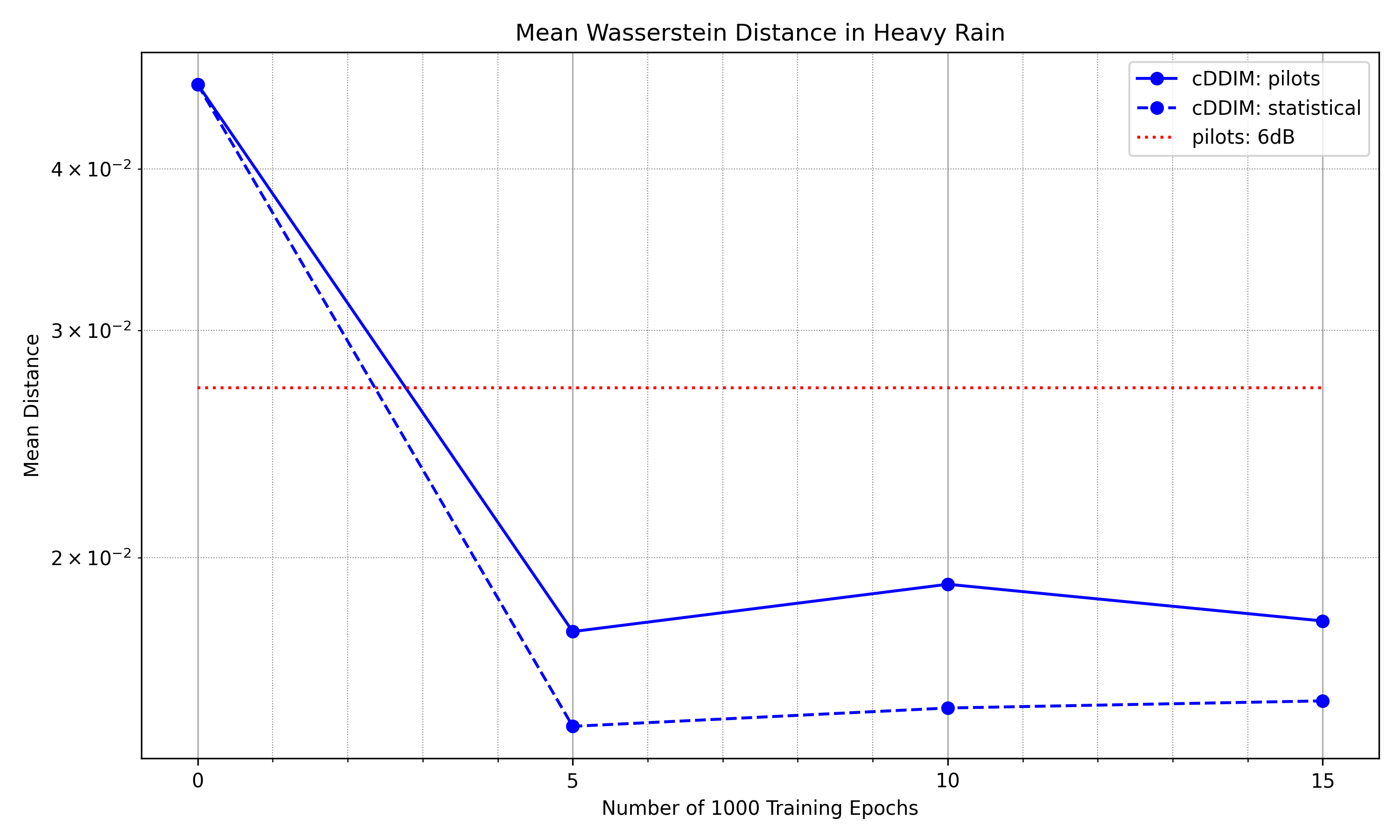}
\centering
 \captionsetup{font=small}
\caption{Wasserstein $p$-distance between generated samples and the test dataset at varying epochs for DMs trained with and without heavy weather samples.}
\label{fig:wass}
\end{figure}
Figure~\ref{fig:wass} illustrates the Wasserstein distance for channels under heavy rain conditions with $p=2$. It can be observed that while both models converge,  the model trained on ground-truth channel samples achieves the minimum distance. This is not surprising because the cDDIM based on \eqref{channel} learns the true distribution of the channel.

However, the main insight is that the Wasserstein distance achieved by the pilot-based channel estimation method is higher than that of both cDDIM models. The Wasserstein distance was calculated using pilot-based channels at a relatively low SNR value of $6\;\rm  dB$. The results indicate that the diffusion-based approach outperforms the conventional pilot-based method in terms of distributional similarity to the true channel.

\subsection{Downstream Task}
To assess the quality and realism of the channels generated by the diffusion model, we benchmark its performance against a conventional pilot-based channel estimation framework. In this section, we assume that neither the transmitter nor the receiver knows the true channel distribution described in \eqref{channel}. Instead, the diffusion model is trained solely using estimates of the channel, which are conventionally assumed to be available. When generating heavy-weather channel realizations using the CDDIM framework, the only required side information is the precipitation rate for rain and snow, or the liquid water density in the case of fog. In contrast, the pilot-based benchmark assumes full CSIT, which further highlights an advantage of the proposed approach. The precoder/decoder structure follows the classical SVD decomposition approach.

To integrate the generative model into this evaluation, we adopt a look-up table approach based on a channel's label. We create a dataset solely from samples generated by the DM, where a sample is the channel and its corresponding label (e.g. heavy rain at 10 mm/h). These samples are generated for each weather condition at the rates given in Table \ref{table:1} for heavy precipitation/density. We assume that the BS has telemetry data on hand, which it uses to find the channel with the label most similar to its real-time assessment. Thus, the BS pulls a channel estimate from a pre-generated sample set instead of a traditional pilot-based approach, which is then passed to calculate the optimal precoder and decoder. This enables us to quantify its utility through the resulting BER and outage performance, allowing us to measure any degradation relative to the baseline estimator and verify the practical relevance of the DM based channels within a learning-driven communication system.

\begin{figure}[h]
\vspace{-4mm}
\includegraphics[width=1\linewidth]{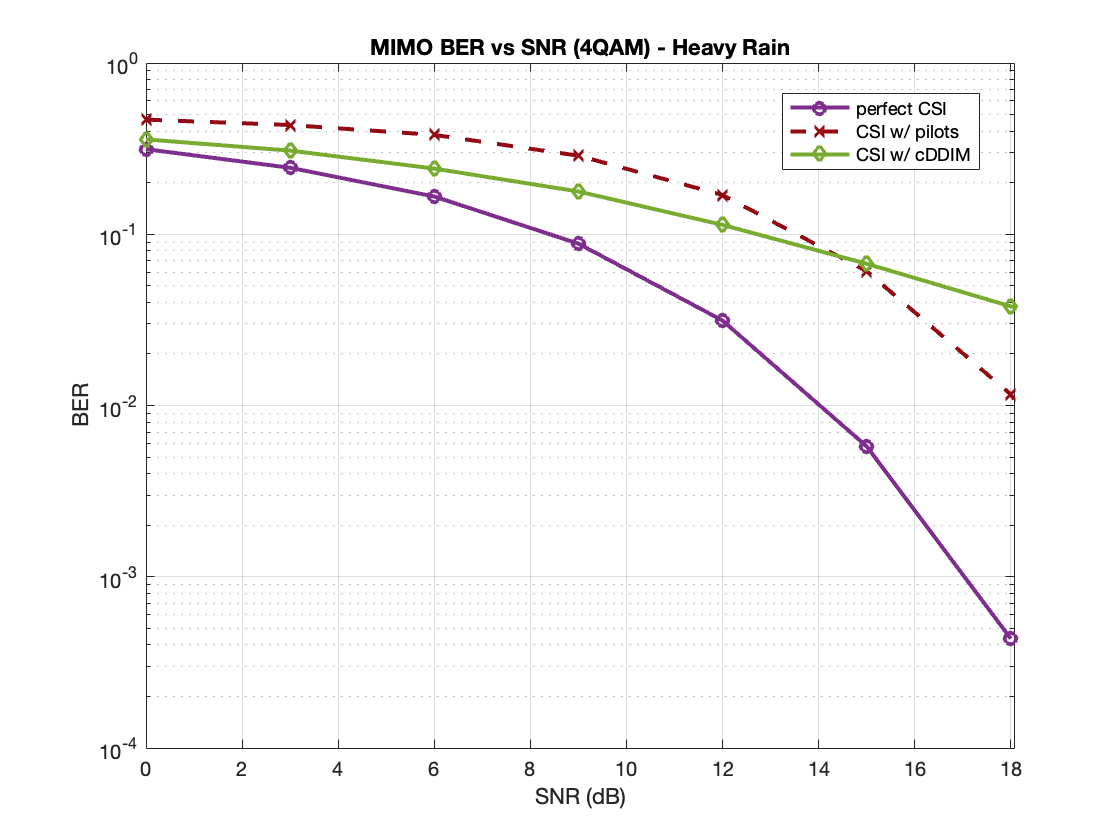}
\centering
 \captionsetup{font=small}
\caption{Uncoded 4QAM BER performance for perfect CSI, pilot-based, and cDDIM based channel estimation in heavy rain.}
\label{fig:uncoded_ber}
\end{figure}
\begin{figure}[h]
\vspace{-5mm}
\includegraphics[width=1\linewidth]{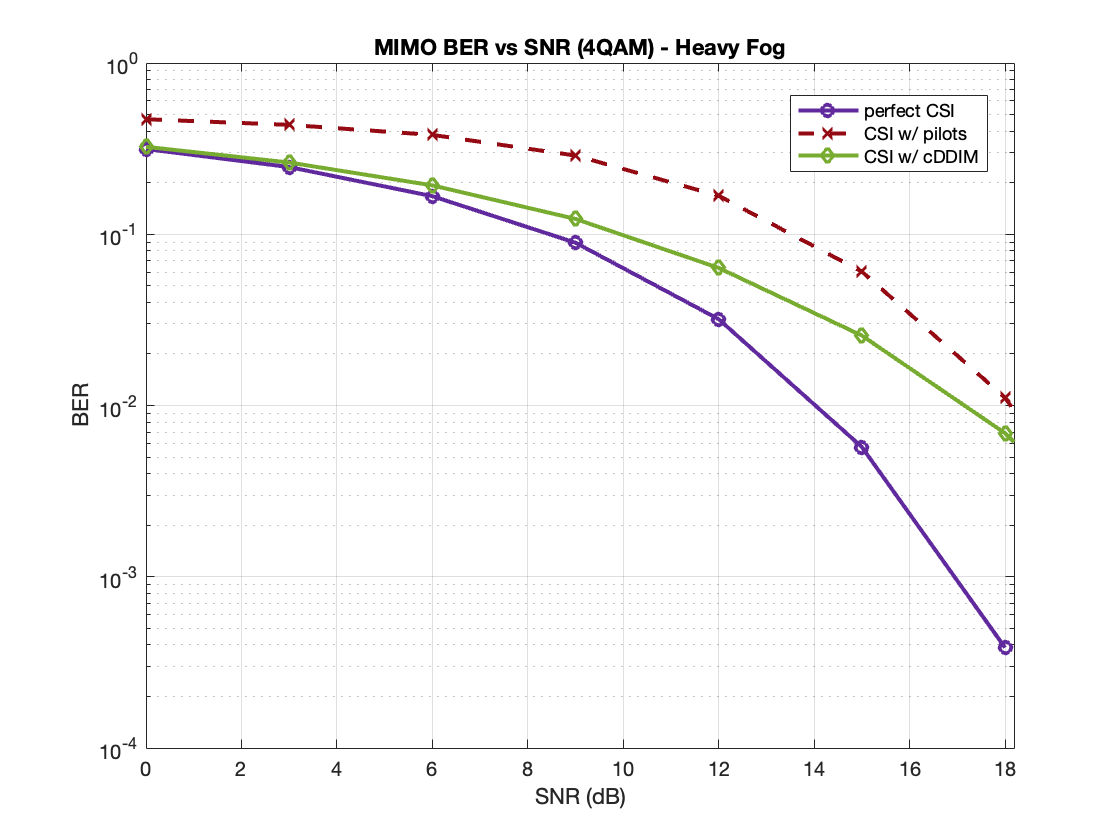}
\centering
 \captionsetup{font=small}
\caption{Uncoded 4QAM BER performance for perfect, pilots, and cDDIM based channel estimation in heavy fog.}
\label{fig:ber_fog}
\end{figure}

Figure~\ref{fig:uncoded_ber} illustrates the BER versus SNR performance under heavy rain, while Figure~\ref{fig:ber_fog} presents the corresponding results for heavy fog, where each point is obtained as a Monte Carlo average over $10^4$ iterations under uncoded 4QAM modulation. In adverse weather conditions, wireless systems are typically expected to operate in the low-SNR regime. It can be clearly observed that at low and moderate SNR values (e.g., $\mathrm{SNR} \leq 10\,\mathrm{dB}$), the channels generated by the diffusion model outperforms the pilot-based channel estimation benchmark and achieves a performance closer to the perfect CSI scenario. 

A crossover behavior is observed as the SNR increases, where the pilot-based approach begins to outperform the generative model. This is expected since pilot-assisted estimation provides increasingly accurate channel observations in the high-SNR regime, whereas the cDDIM framework predicts channel realizations based on the weather condition. Furthermore, the heavy fog scenario remains much closer to the perfect CSI benchmark compared to heavy rain. This can be attributed to the fact that heavy rain exhibits a wider unexplored operating range and generally induces more severe scattering and path-loss effects than fog. Consequently, the crossover occurs at approximately $20\,\mathrm{dB}$ for heavy fog, whereas it occurs near $15\,\mathrm{dB}$ for heavy rain.

\begin{figure}[btph]
\vspace{-4mm}
\includegraphics[width=1\linewidth]{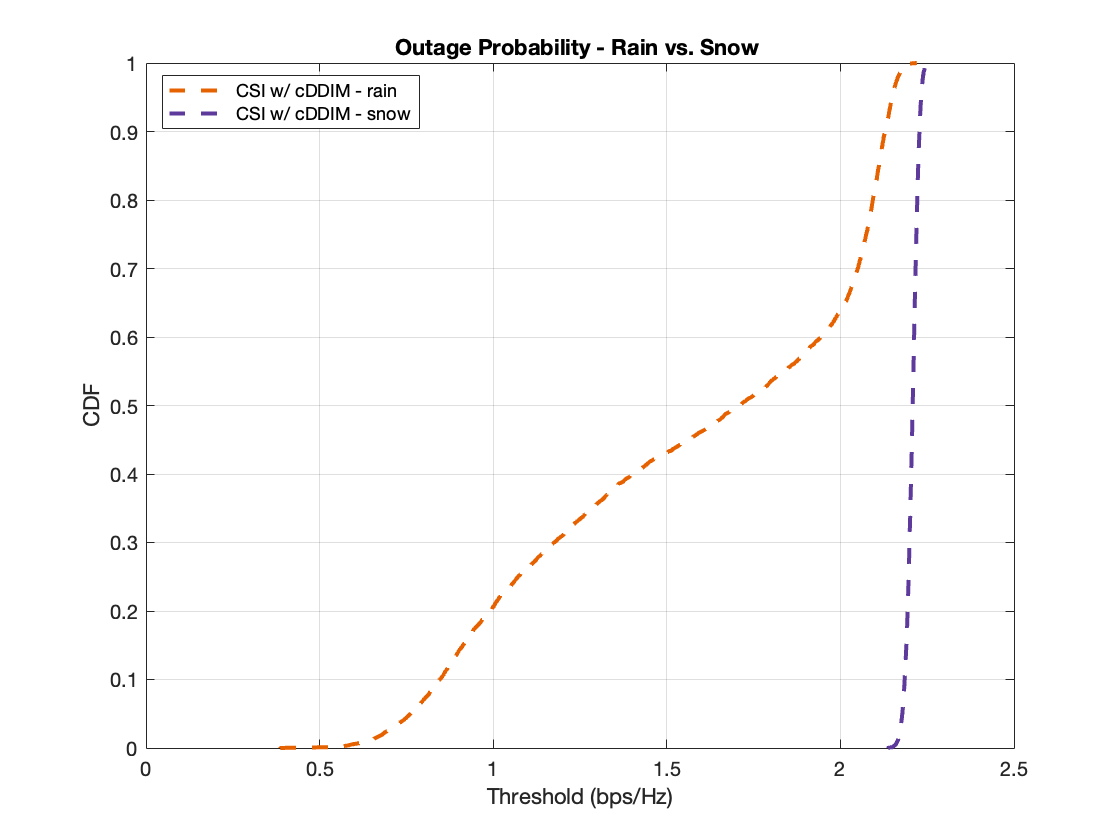}
\centering
 \captionsetup{font=small}
\caption{Outage probability given a spectral efficiency (SE) threshold for cDDIM-based CSI in rain and snow.}
\label{fig:outage}
\end{figure}

In Figure~\ref{fig:outage}, the outage probabilities in heavy rain and snow conditions of our approach are plotted. We note that the outage probability in rainy conditions is worse than in snow in low SNR. In the case of rain, the diffusion model is trained on a dataset whose parameter range is significantly narrower than that encountered during testing (the precipitation levels for heavy rain span a much larger range than that of snow), leading to degraded performance. For snow, the precipitation range is more limited, which explains how the DM is able to produce statistically-similar channels across all snow intensities. Future work could include a wider range of samples while training the model, such as channels experiencing heavy fog and rain, or altering the weather-impaired model for specific environments.

\section{Conclusion}
This work explored the use of generative diffusion models to synthesize wireless channel samples under adverse weather conditions. By conditioning on rain, snow, and fog intensity, the model aims to capture realistic variations in CSI without relying on extensive and often impractical data collection. Instead, it leverages measurements obtained under milder weather conditions, where channel observations are more reliable and can be trusted. We employed a stochastic channel simulator to generate baseline channels and adopted an existing cDDIM architecture to produce samples that would be unreliable if acquired using conventional methods. The generated channels were evaluated both statistically, using the Wasserstein distance, and functionally, through their impact on downstream channel estimation tasks measured via BER and outage probability.

Our analysis of the Wasserstein distance demonstrates that diffusion models can effectively map channels across weather conditions using only weather intensity information. This suggests that severe weather channel impairments may be captured using coarse environmental measurements rather than full real-world channel datasets. 

In terms of practical performance advantages, the use of diffusion models eliminates the need for extensive pilot overhead and requires only measurements of weather intensity. Furthermore, channels based on a diffusion model will be agnostic to channel aging. In contrast, even with perfect knowledge of the ground-truth distribution, a significantly larger number of parameters would be required to accurately generate the channel, which further demonstrates the scalability of our method.

Finally, our downstream evaluations highlight the practical benefits of diffusion based CSI acquisition. At low and moderate SNR, the regimes expected under adverse weather, we observe measurable gains in both BER performance and outage probability relative to pilot-based estimation. These results indicate that diffusion models can improve performance and expand feasibility in challenging communication environments. 

\section{Acknowledgment}
The authors sincerely thank Jeffrey G. Andrews for his insightful discussions and valuable guidance, which significantly contributed to the development of this work.
This work is supported by the National Defense Science Engineering and Graduate (NDSEG) fellowship. Any opinions, findings, conclusions or recommendations expressed in this material are those of the authors and do not necessarily reflect the views of NDSEG.

\bibliographystyle{IEEEtran}
\bibliography{bib.bib}

\end{document}